\documentclass[pra,twocolumn, notitlepage,superscriptaddress, nofootinbib]{revtex4-1}

\usepackage{fancyhdr}
\fancyheadoffset[L,R]{\headrulewidth}
\renewcommand{\headrulewidth}{0.6pt}

\usepackage{cancel}
\usepackage[normalem]{ulem}

\usepackage{cancel}

\usepackage{graphicx}
\usepackage{tikz,pgf}
\usepackage{color}
\usepackage{array}
\usepackage{xcolor}
\usepackage[dvips]{epsfig}
\usepackage{epsfig}
\usepackage{enumerate}
\usepackage[latin1]{inputenc}
\usepackage[T1]{fontenc}
\usepackage[hang]{subfigure}
\usepackage{bbm, dsfont}
\usepackage{amsfonts,amsmath,amssymb,stmaryrd}
\usepackage{ulem}

\usepackage{bbold}

\usepackage{simplewick}

\newcommand{\bra}[1]{\langle #1 | \,}
\newcommand{\ket}[1]{\, | #1 \rangle}

\newcommand{\ga}{\ga}

\newcommand{\bl}{\begin{linenomath*}}
\newcommand{\el}{\end{linenomath*}}

\newcommand{\bea}{\begin{eqnarray}}
\newcommand{\eea}{\end{eqnarray}}

\renewcommand{\ga}{\hat\gamma}

\definecolor{dgreen}{rgb}{0.0, 0.5, 0.0}

\usepackage{hyperref}

\begin{document}

%%%%%%%%%%%%%%%%%%%%%%%%%%%%%%%%%%%%%%%%%%%%%%%%%%
\title{Angular self-localization of impurities  rotating in a bosonic bath}
%%%%%%%%%%%%%%%%%%%%%%%%%%%%%%%%%%%%%%%%%%%%%%%%%%

\author{Xiang Li}
\affiliation{IST Austria (Institute of Science and Technology Austria), Am Campus 1, 3400 Klosterneuburg, Austria}

\author{Robert Seiringer}
\affiliation{IST Austria (Institute of Science and Technology Austria), Am Campus 1, 3400 Klosterneuburg, Austria}

\author{Mikhail Lemeshko}
\email{mikhail.lemeshko@ist.ac.at}
\affiliation{IST Austria (Institute of Science and Technology Austria), Am Campus 1, 3400 Klosterneuburg, Austria}

\begin{abstract}

The existence of a self-localization transition in the polaron problem has been under an active debate ever since Landau suggested it 84 years ago. Here we reveal the self-localization transition for the rotational analogue of the polaron -- the angulon quasiparticle. We show that, unlike for the polarons, self-localization of angulons occurs at finite impurity-bath coupling already at the mean-field level. The transition is accompanied by the spherical-symmetry breaking of the angulon ground state and a discontinuity in the first derivative of the ground-state energy. Moreover, the type of symmetry breaking is dictated by the symmetry of the microscopic impurity-bath interaction, which leads to a number of distinct self-localized states. The predicted effects can potentially be addressed in experiments on cold molecules trapped in superfluid helium droplets and ultracold quantum gases, as well as on electronic excitations in solids and Bose-Einstein condensates.

\end{abstract}

\maketitle

\section{Introduction}

In 1933, Landau predicted that electrons moving in solids can undergo a self-localization transition~\cite{first_polaron}. The latter takes place when the electron-induced distortion of the crystal lattice is strong enough to affect the motion of the electron itself, confining its wavefunction in space. This seminal work led to the concept of the polaron -- a quasiparticle composed of  a free electron `dressed' by a field of lattice vibrations~\cite{Pekar47, Pekar48, LandauPekar48}. By now polarons emerged as a key theoretical tool to describe electron transport in condensed matter physics and chemistry~\cite{AppelPolarons, EminPolarons, PolaronsExcitons, Devreese15}, as well as to understand the behaviour of atomic impurities in ultracold quantum gases~\cite{ChikkaturPRL00, SchirotzekPRL09, PalzerPRL09, KohstallNature12, KoschorreckNature12, SpethmannPRL12, FukuharaNatPhys13, ScellePRL13, Cetina15, MassignanRPP14,Jorgensen2016, Hu16, Cetina2016}.

While the phenomenon of self-localization attracted attention of several generations of physicists, its existence in various polaron models is still under an active debate.  In the early 1950's, Fr\"ohlich introduced a purely microscopic model describing interaction of a point charge with optical phonons~\cite{frohlich_hamiltonian, Devreese15}. Several theories predicted the existence of a self-localization transition in Fr\"ohlich polaron~\cite{Gaussian_model1, L_M_theory1, Manka_approximation, L_M_theory2, Gaussian_model2} as well as in related models such as a polaronic exciton \cite{self_trapping_in_exciton}  and an optical polaron in an external magnetic~\cite{self_trapping_in_magnetic_field} or Coulomb  potential~\cite{self_trapping_in_Coulomb_potential}. Later, however, the effect had been proven to be a consequence of the approximations employed rather than an intrinsic property of the polaron Hamiltonian~\cite{FisherPRB86, no_self_trapping1, no_self_trapping2, no_self_trapping_numerical, Feranchuk05}. Self-localization has also been predicted for a particle coupled to acoustic phonons and collective excitations in a Bose-Einstein condensate (BEC)~\cite{acoustic_polaron2, acoustic_polaron3, acoustic_polaron1, self_trapping_in_acoustic_polaron_and_BEC_numerical1, self_trapping_in_BEC1, self_trapping_in_BEC2, BEC_impurity_soliton, BEC_impurity_attractive_potential, BEC_impurity_Feynman_approach, BEC_impurity_multiple_impurities1, BEC_impurity_multiple_impurities2, BEC_impurity_reduce_dimention, self_trapping_in_acoustic_polaron_and_BEC_numerical3, BEC_impurity_finite_temperature, BEC_impurity_coupled_to_laser_cavity, self_trapping_in_acoustic_polaron_and_BEC_numerical4}. Still, recent numerical results  reveal a smooth crossover of the BEC-polaron energy between the weakly- and strongly-coupled regimes, suggesting an absence of the self-localization transition~\cite{BEC_impurity_renormalization_group_approach1,BEC_impurity_renormalization_group_approach2,BEC_impurity_renormalization_group_approach3,BEC_impurity_renormalization_group_approach4,BEC_impurity_renormalization_group_approach5, ArdilaPRA15}.

In this paper, we introduce a novel platform to study the self-localization/delocalization transition -- that consisting of an impurity with nonzero orbital angular momentum interacting with a bosonic bath. Such a setting can represent e.g.\ an ultracold alkaline or alkaline-earth dimer immersed into a BEC~\cite{cold_molecule_book1, JinYeCRev12}, a polyatomic molecule trapped inside a superfluid helium nanodroplet~\cite{molecule_experiment_angular_superfluid_helium}, an electron bubble in liquid helium~\cite{TempereEPJB, VadakkumbattNatComm14}, as well as an electronic excitation in a BEC~\cite{BalewskiNature13} or a solid~\cite{Mahan90, WeissBook}.

Recently it has been shown that the behavior of such orbital impurities in a bosonic bath can be described in terms of the \textit{angulon} quasiparticle. The angulon consists of a quantum rotor surrounded by a many-particle field of boson excitations, and thereby represents the rotational counterpart of the polaron quasiparticle. However, the non-Abelian algebra and discrete energy spectrum associated with quantum rotations makes the angulon physics remarkably different from that of polarons~\cite{angulon1, angulon2, LemSchmidtChapter, Bikash16, Redchenko16, Yakaboylu17}.

 Introducing angulons allows to substantially simplify problems involving angular momentum exchange between an impurity and a many-particle bath. For instance, interaction of rotating molecules with superfluid helium received a great deal of attention in the past, both from the experimental and theoretical perspectives~\cite{molecule_experiment_angular_superfluid_helium, SzalewiczIRPC08, Babichenko99}. Most of theoretical studies, however, relied on extensive numerical computations for finite-size helium clusters, based on path-integral, variational, and  diffusion quantum Monte-Carlo techniques~\cite{BarnettHeSF6, BlumeJCP96, LeeHeSF6, KwonJCP96, KwonJCP00, PaesaniJCP01, MoroniPRL03, PatelJCP03, TangPRL04, PaesaniJCP04, MoroniJCP04,  ZillichHeC2H2, ZillichHeHCN,  PaesaniPRL05, ZillichJCP05, PaoliniJCP05, TopicJCP06,   VielJCP07,SkrbicJPCA07, MiuraJCP07, vonHaeftenHeCO, ZillichHeLiH, MarkovskiyJPCA09, RodriguesIRPC16}. As one of us has recently demonstrated, the complex many-particle problem can be drastically simplified if one assumes that molecules in helium form angulons~\cite{LemeshkoDroplets16}. The angulon theory furnishes the effective molecular moments of inertia in closed form and provides them with a transparent physical interpretation. Furthermore, good agreement of the angulon theory with experiment provides strong evidence that angulons are indeed formed out of molecules immersed in superfluid $^4$He~\cite{LemeshkoDroplets16}. Finally, the simplicity of the angulon theory allows to study dynamical properties of many-particle systems which are extremely challenging to address using other techniques. As an example, the theory is able to reproduce non-adiabatic molecular dynamics in helium nanodroplets following a short laser pulse, as observed in experiment~\cite{Shepperson16}.

 In what follows, we develop a strong-coupling angulon theory based on the mean field approximation. Our main goal is to reveal the possibility of the angular self-localization transition, corresponding to the breaking of the spherical symmetry, and compare the results of the angulon and polaron models. We demonstrate that -- within the mean-field approximation -- such a transition takes place at a finite impurity-bath coupling strength and is accompanied by a discontinuity in the first derivative of the  angulon ground-state energy. Furthermore, we demonstrate that the type of the symmetry breaking depends on the symmetry of the microscopic impurity-atom potential, which results in a number of distinct self-localized states. It is important to note that angulon self-localization takes place in the continuous space of the impurity angles, and is therefore fundamentally different  from localization in the Caldeira-Leggett and related models~\cite{LeggettRMP87, WeissBook}.

\section{Mean-field approach to angular self-localization}

In Ref.~\cite{angulon1} it was shown that the interaction of a rotating impurity with a bosonic bath can be described using the following Hamiltonian:
\begin{multline}
\hat H = B\hat{\bf{J}}^2 + \sum_{k\lambda\mu}\omega_k\hat{b}^+_{k\lambda\mu}\hat{b}_{k\lambda\mu} \\
+ \sum_{k\lambda\mu}U_{\lambda}(k)\left[Y^*_{\lambda\mu}(\hat{\theta},\hat{\phi})\hat{b}^+_{k\lambda\mu} + Y_{\lambda\mu}(\hat{\theta},\hat{\phi})\hat{b}_{k\lambda\mu}\right]
\label{eq:1}
\end{multline}
where $\sum_{k}\equiv\int dk$ and $\hbar\equiv1$. In what follows, Eq.~\eqref{eq:1} will be referred to as the `angulon Hamiltonian.' For simplicity, we consider a linear-rotor impurity whose kinetic energy is given by the first term of Eq. (\ref{eq:1}), where $B$ is the rotational constant and $\hat{\bf{J}}$ is the angular momentum operator. The rotor eigenstates, $\vert j,m\rangle$, are labeled by the angular momentum, $j$, and its projection, $m$,  on the laboratory $z$-axis.
While  a rigid linear rotor provides a perfect model for rotation of a diatomic  molecule, the first term of Eq.~\eqref{eq:1} is straightforward to extend  to more complex polyatomic species or electronic states with nonzero angular momentum~\cite{BernathBook, LevebvreBrionField2, RudzikasAtomicSpec}. The second term of Eq. (\ref{eq:1}) is the kinetic energy of the bosons with a dispersion relation $\omega_k$. For convenience, the bosonic creation and annihilation operators, $\hat{b}^+_{\bf{k}}$ and $\hat{b}_{\bf{k}}$,  are expressed in the angular momentum basis:
\begin{equation}
\hat{b}^+_{k\lambda\mu}=k(2\pi)^{-3/2}\int d\Omega_k\hat{b}^+_\mathbf{k}i^\lambda Y^*_{\lambda\mu}(\Omega_k)
\end{equation}
Here, $k=|\bf{k}|$, while $\lambda$ and $\mu$ define the boson angular momentum and its projection onto the laboratory-frame $z$-axis, see Refs.~\cite{angulon1, angulon2, LemSchmidtChapter} for details. The third term of Eq.~\eqref{eq:1} describes the impurity-bath interaction which explicitly depends on the impurity orientation in the laboratory frame, as given by the spherical harmonic operators, $Y_{\lambda\mu}(\hat{\theta},\hat{\phi})$~\cite{Varshalovich}. The coupling constants,  $U_{\lambda}(k)$, parametrize the interaction of the impurity with phonons carrying angular momentum $\lambda$ and linear momentum $k$. In Ref.~\cite{angulon1}  analytic expressions for $\omega_k$ and $U_{\lambda}(k)$ were provided for the case of an ultracold molecule rotating inside a weakly-interacting BEC. For more involved cases, such as molecules in superfluid helium or electronic excitations in solids, the corresponding coupling constants can be used as phenomenological parameters. However, as we demonstrate below, the qualitative properties of the self-localization transition do not depend on the momentum dependence of the impurity-bath interaction and are determined solely by its symmetry. Therefore, in what follows we will consider the Hamiltonian~\eqref{eq:1} from a completely general perspective, without focusing on a particular physical system.

Here we use a mean-field theory analogous to the Landau-Pekar approximation previously used to describe polarons~\cite{first_polaron,Pekar_ansatz, Devreese15}. Namely, we approximate the angulon state by the product ansatz,
\begin{equation}
|\psi\rangle = \vert \text{imp}  \rangle \otimes \vert \text{bos} \rangle,
\label{eq:2}
\end{equation}
where $\vert \text{imp}   \rangle$ and $\vert \text{bos} \rangle$ describe the impurity and the bosonic bath, respectively. Although  Eq.~\eqref{eq:2} corresponds to a low-level approximation to the solutions of the angulon Hamiltonian~\eqref{eq:1}, already at this level  the angulons feature quite a rich physical behavior, substantially different from that of polarons, as we shall demonstrate below. Furthermore, while the validity of the mean field approximation to study the localization transition is debatable~\cite{no_self_trapping2}, it has been shown to provide a good estimate for the ground-state energy in the strong-coupling limit~\cite{Donsker83, Lieb97}.

Within the mean-field approach, one can write an effective bosonic Hamiltonian, $\hat H_B$, for every state of impurity $\vert \text{imp}  \rangle$ separately:
\begin{multline}
\hat H_B \equiv \bra{\text{imp}} \hat H \ket{\text{imp}} = B \langle  \text{imp} \vert \hat{\mathbf J}^2 \vert \text{imp}\rangle + \sum_{k\lambda\mu}\omega_k\hat{b}^+_{k\lambda\mu}\hat{b}_{k\lambda\mu} \\
+ \sum_{k\lambda\mu}U_{\lambda}(k)\left[ \langle  \text{imp} \vert Y^*_{\lambda\mu}(\hat{\theta},\hat{\phi}) \vert \text{imp}\rangle \hat{b}^+_{k\lambda\mu} +  \text{h.c.}\right]
\label{eq:HB}
\end{multline}
The ground state of Eq.~\eqref{eq:HB}, in turn, can be obtained analytically as a bosonic coherent state:
\begin{equation}
\vert \text{bos} \rangle = \exp\left[\sum_{k\lambda\mu}\left(\beta^{\text{imp}}_{k\lambda\mu}~\hat{b}_{k\lambda\mu} -   \beta^{\text{imp} \ast}_{k\lambda\mu}~\hat{b}^+_{k\lambda\mu}\right)\right] \vert 0 \rangle
\label{eq:6}
\end{equation}
where $\vert 0 \rangle$ is the bosonic vacuum, and the coefficients
\begin{equation}
\beta^\text{imp}_{k\lambda\mu} = \frac{U_{\lambda}(k)}{\omega_k} \langle   \text{imp}  \vert Y_{\lambda\mu}(\hat{\theta},\hat{\phi})\vert   \text{imp}  \rangle
\end{equation}
parametrically depend on the state of the impurity $\vert \text{imp}  \rangle$.

\begin{figure}[th!]
\centering
\includegraphics[width=0.45\textwidth]{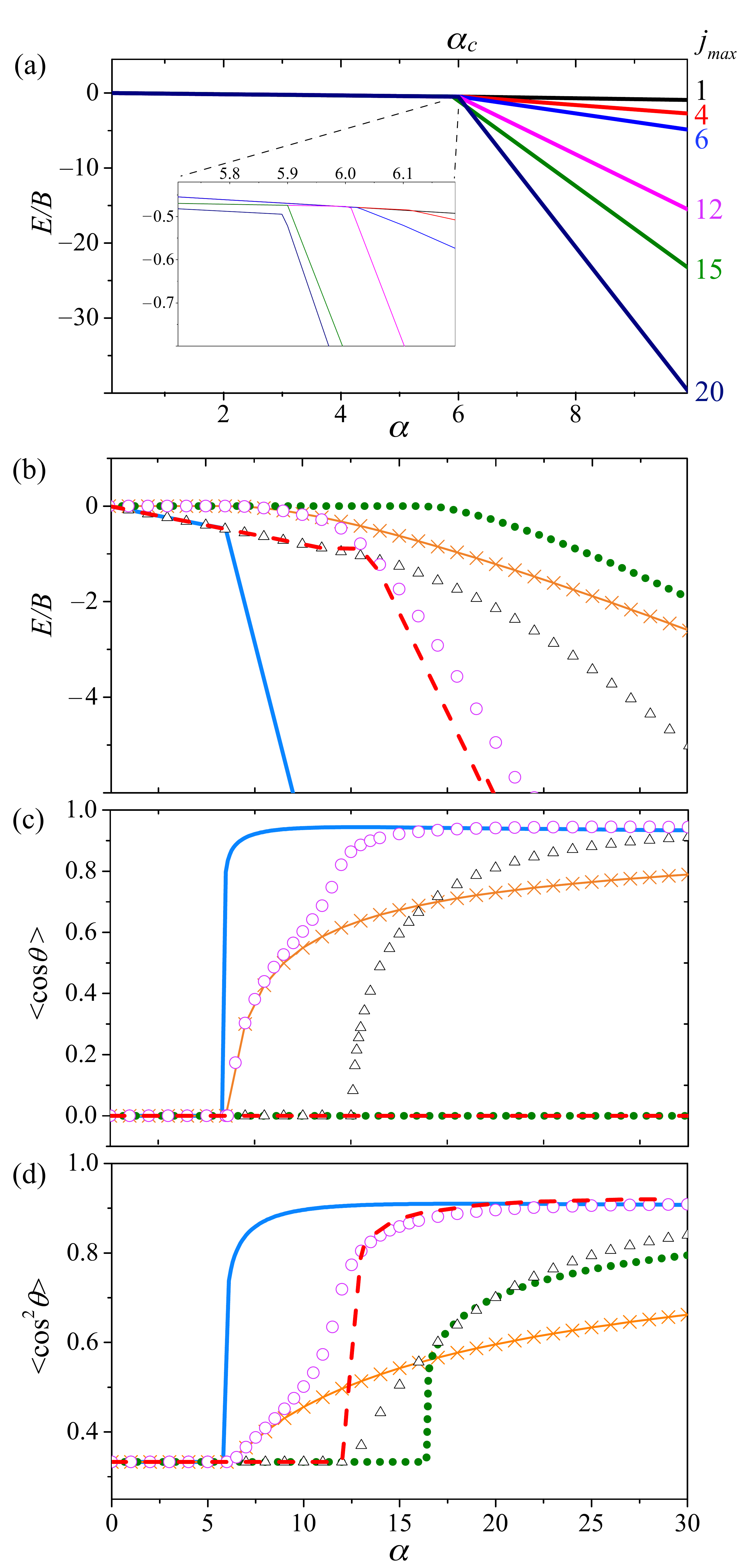}
\caption{ (Color online) (a) Dependence of the impurity ground state energy on the magnitude of the constant impurity-bath coupling strength, $\alpha_{\forall \lambda} =\alpha$, at different values of the cutoff $j_\text{max}$. The inset shows the vicinity of the transition point. (b) The case of $j_\text{max} = 6$ for various types of the impurity-bath interaction: $\alpha_{\forall \lambda} = \alpha$ (blue solid line); \{$\alpha_1 = \alpha$, $\alpha_{\neq 1}  = 0$\} (yellow crosses); \{$\alpha_2 = \alpha$, $\alpha_{\neq 2} =  0$\} (green dotted line); \{$\alpha_\text{odd} = \alpha$, $\alpha_\text{even} =  0$\} (purple circles); \{$\alpha_\text{even} = \alpha$, $\alpha_\text{odd} =  0$\}  (red dashed line); as well as $\alpha_\lambda = \alpha/(1+\lambda)$ (empty triangles). (c) Same as in (b), but for the orientation cosine of the impurity. (c) Same as in (b), but for the alignment cosine of the impurity.}
\label{fig:1}
\end{figure}

The corresponding angulon ground-state  energy is given by (in units of B):
\begin{equation}
\frac{E}{B} = \langle  \text{imp} \vert \hat{\mathbf J}^2 \vert  \text{imp} \rangle -  \sum_{\lambda\mu} \alpha_\lambda  \vert \langle   \text{imp}  \vert Y_{\lambda\mu}(\hat{\theta},\hat{\phi})\vert   \text{imp}  \rangle\vert^2,
\label{eq:energy}
\end{equation}
where we introduced  dimensionless impurity-bath interaction parameters:
\begin{equation}
\label{eq:alpha}
\alpha_\lambda = \sum_k \frac{U^2_\lambda(k)}{\omega_k B}
\end{equation}
 From the form of Eq.~\eqref{eq:energy}, one can already see that the ground state energy depends on the momentum distribution of the impurity-bath coupling only through the parameters $\alpha_\lambda$.

Note that, irrespectively of the sign of the coupling $U_\lambda(k)$, the \textit{effective} many-body interaction given by Eq.~\eqref{eq:alpha} is always attractive. This is a general property of Hamiltonians with a linear coupling, such as Eq.~\eqref{eq:1}, or Hamiltonians describing Fr\"ohlich polarons~\cite{Devreese15} or AC Stark effect~\cite{ScullyZubairy, LemKreDoyKais13}. Effective repulsive interactions, on the other hand, take place e.g.\ for the repulsive polaron branch observed in ultracold gases~\cite{Rath2013, Cetina2016} or for alkali atoms and dimers trapped on the surface of helium nanodroplets~\cite{MudrichIRPC14}. Describing such systems requires  nonlinear terms in the impurity-boson interaction of Eq.~\eqref{eq:1}, and therefore goes beyond the effective model studied in this paper.

In what follows, we solve Eq.~\eqref{eq:energy} variationally assuming a general form of the impurity state:
\begin{equation}
\vert \text{imp} \rangle = \sum_{j=0}^{j_\text{max}} \sum_{m=-j}^{j} c_{jm} \vert jm\rangle,
\label{eq:10}
\end{equation}
where $c_{jm}$ are the variational parameters obeying a normalization condition,
\begin{equation}
\sum_{jm} \vert c_{jm} \vert^2 \equiv 1,
\end{equation}
and $j_\text{max}$ provides an angular momentum cutoff.
In order to simplify the variational calculations, we assume rotational symmetry with respect to the $z$-axis and therefore restrict the variational space to the $m=0$ subspace. Such a restriction does not amount to an approximation when the interaction is purely attractive, in the sense that $\sum_{\lambda,\mu} \alpha_\lambda Y_{\lambda,\mu}(\mathbf{\Omega}) Y^*_{\lambda,\mu}(\mathbf{\Omega}') $ is an increasing function of $\mathbf{\Omega\cdot \Omega}'$~\cite{Baernstein76}. For all interaction potentials considered here, the restriction to the $m=0$ subspace does not affect the minimizers.

\section{Cases of different symmetry}

Let us start by considering the most transparent model  with the coupling constants independent of angular momentum, $\alpha_{\forall \lambda} = \alpha$. Fig.~\ref{fig:1}(a) shows the dependence of the ground state energy on the magnitude  of the constant potential,  $\alpha$,  for different values of the cutoff $j_\text{max}$. We see that the energy possesses a nonanalyticity around the critical value $\alpha_c \sim 6$. The inset of Fig.~\ref{fig:1}(a) zooms into the vicinity of the non-analyticity point: one can see that it slightly shifts to the left the larger $j_\text{max}$ is. In the limit of $j_\text{max} \to \infty$, the critical point approaches $\alpha_c \approx 5.85$. In this case, the minimizer for $\alpha<\alpha_c$ corresponds to a spherically-symmetric ground state ($c_j = \delta_{j0}$), while for $\alpha>\alpha_c$ the ground state corresponds to a $\delta$-function in the angular space (all $c_j=j_\text{max}^{-1/2}$). This result coincides with the one obtained by solving a corresponding nonlinear Schr\"odinger equation on a two-dimensional plane, where $\alpha_c \approx 1.86225 \pi = 5.85043 $ was found to be the critical coupling constant~\cite{WeinsteinCommMathPhys83}. Since the qualitative behavior of the ground-state energies does not  depend  on the cutoff, in what follows we focus on the case of $j_\text{max} = 6$.

\begin{figure}[t]
\centering
\includegraphics[width=0.45\textwidth]{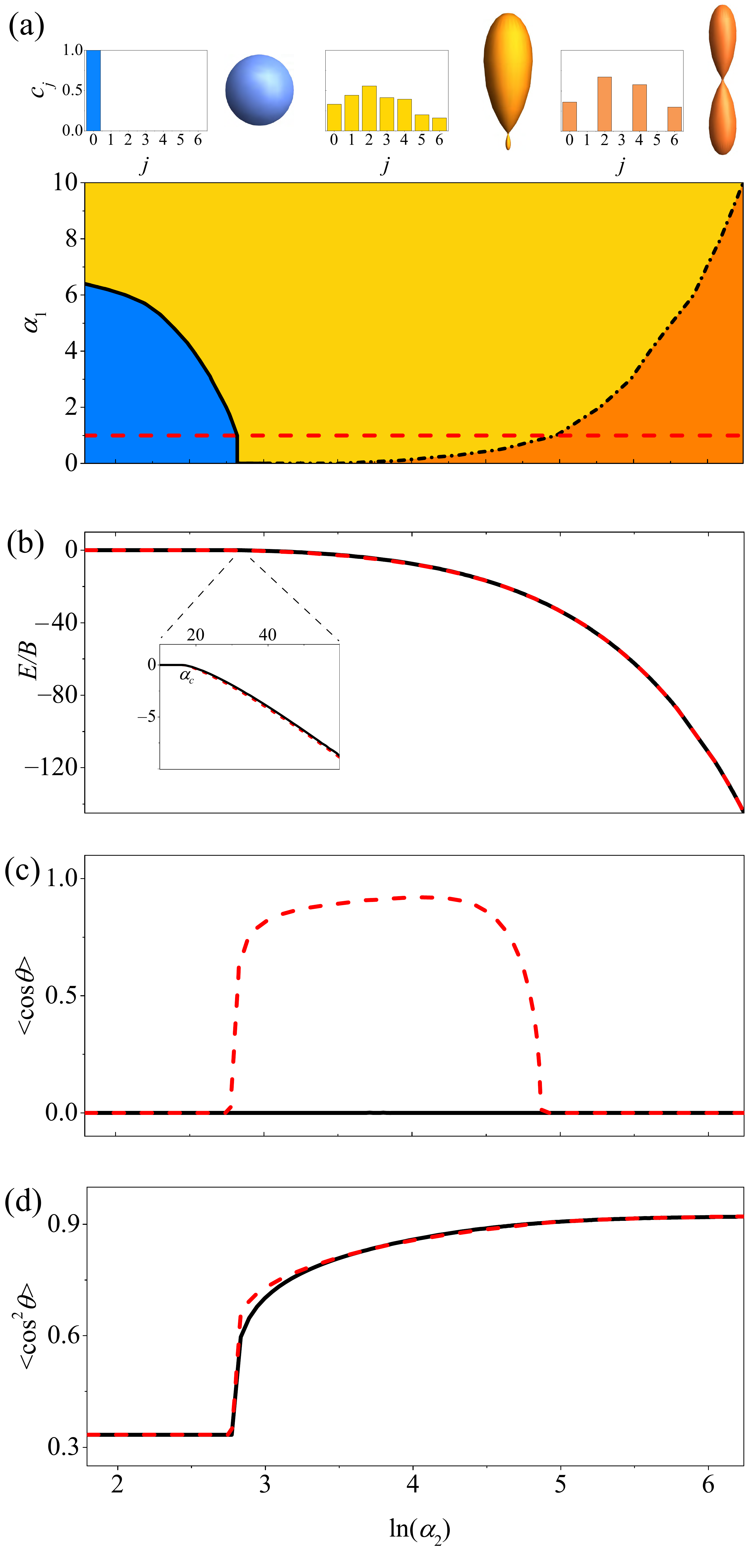}
\caption{ (Color online) (a) `Localization diagram' for the angulon ground state, depending on the magnitudes of $\alpha_1$ and $\text{ln}(\alpha_2)$  (with other $\alpha_\lambda$ set to zero). (b) Dependence of the ground-state energy on $\text{ln}(\alpha_2)$ for $\alpha_1 = 0$ (black solid line) and $\alpha_1 = 1$ (red dashed line). The inset shows the vicinity of the localization transition. (c) Same as in (b), but for the orientation cosine of the impurity. (c) Same as in (b), but for the alignment cosine of the impurity.}
\label{fig:2}
\end{figure}

For most systems available in experiment, the impurity-bath interaction is dominated by the first few $\lambda$-terms and is usually decaying such that $\alpha_\lambda$ is negligibly small for $\lambda \gtrsim 5$~\cite{StoneBook13, Bikash16}. In order to cover experimentally relevant cases and illustrate the fact that the transition is universal, we consider several different types of the impurity-bath interactions. Fig.~\ref{fig:1}(b) shows the behavior of ground-state energies for the cases of  \{$\alpha_1 = \alpha$, $\alpha_{\neq 1}  = 0$\} (yellow crosses); \{$\alpha_2 = \alpha$, $\alpha_{\neq 2} =  0$\} (green dotted line); \{$\alpha_\text{odd} = \alpha$, $\alpha_\text{even} =  0$\} (purple circles); \{$\alpha_\text{even} = \alpha$, $\alpha_\text{odd} =  0$\} (red dashed line); as well as $\alpha_\lambda = \alpha/(1+\lambda)$ (empty triangles). For comparison, the case of  $\alpha_{\forall \lambda} = \alpha$ is shown by the blue solid line. One can see that while the position of the transition point shifts depending on the form of the interaction, the transition still takes place independently of the latter.

In order to get insight into the angular symmetry of the localized impurity, in Figs.~\ref{fig:1}(c) and (d) we plot the orientation cosines,
\begin{equation}
\langle \cos \theta \rangle \equiv \bra{ \text{imp} }\cos \hat \theta \ket{ \text{imp} },
\end{equation}
and the alignment cosines,
\begin{equation}
\langle \cos^2 \theta \rangle \equiv \bra{ \text{imp}} \cos^2 \hat \theta \ket{ \text{imp} },
\end{equation}
of the impurity.
One can see that to the left of the transition point, $\langle \cos \theta \rangle = 0$ and $\langle \cos^2 \theta \rangle = 1/3$, which reflects the spherical symmetry of the ground angulon state, i.e.\ $c_{jm} = \delta_{j,0}$ in Eq.~\eqref{eq:10}. The transition, on the other hand, is accompanied by an abrupt change in the alignment and/or orientation cosine, which implies the breaking of the impurity spherical symmetry, i.e. angular localization of the angulon. It is important to emphasize that such a symmetry breaking  takes place at a finite value of $\alpha_c$, which is clearly distinct from the case of polarons. There, the same level of approximation -- the Landau-Pekar ansatz~\cite{Devreese15} -- results in a localized impurity already at infinitely weak coupling.  

While the transition occurs independently of the exact form of the $\alpha_\lambda$ distribution, different symmetries of the impurity-bath interaction result in different symmetries of the localized states.  In particular, an interaction dominated by even $\lambda$-terms results in aligned states of definite parity. Such states are characterized by $\langle \cos \theta  \rangle = 0 $ and $\langle \cos^2 \theta \rangle > 1/3$, which is implied by $c_\text{even} \neq 0$ and $c_\text{odd} = 0$ in Eq.~\eqref{eq:10}. On the other hand, the $\alpha_\text{odd}$ terms break the parity symmetry, leading to the oriented localized states with both even and odd $c_j$'s populated. We would like to point out an additional `kink' in  $\langle \cos \theta \rangle$ and  $\langle \cos^2 \theta \rangle$ taking place for \{$\alpha_\text{odd} = \alpha$, $\alpha_\text{even} =  0$\} around $\alpha=10.5$. For the impurity-bath interactions of such a symmetry, the $\Delta j =1$ transitions between the impurity states are dominant. As a result, in the region of $6 \lesssim \alpha \lesssim 10.5$, the impurity is localized mostly in a superposition of the $j=0$ and $j=1$ states.  Around the kink point, however, the magnitude of $c_1$ becomes comparable to the one of $c_0$, which leads to a substantial population of the $j\geq 2$ state and, consequently, stronger localization. However, this point does not correspond to any additional symmetry breaking as the impurity is localized in the oriented state for $\alpha \gtrsim 6$.

Unlike in polarons, the angulon Hamiltonian can feature competing interactions of different symmetry, which results in a richer localization behavior. In order to illustrate the latter, in Fig.~\ref{fig:2}(a) we show the `localization diagram' describing the symmetry of the impurity depending on the magnitudes of $\alpha_1$ and $\alpha_2$ (with other coupling constants set to zero). While $\alpha_{1,2}$ are the interaction components featured by typical molecule-atom potentials~\cite{StoneBook13}, a similar diagram takes place for  $\alpha_\text{even}$ and $\alpha_\text{odd}$ taken as parameters. The top subpanel illustrates the distribution of the $c_j$ coefficients and the wavefunctions for corresponding states. The blue region corresponds to a delocalized, spherically-symmetric ground state with $c_j=\delta_{j0}$, the red one corresponds to an aligned impurity state with only even $c_j$'s populated, while the yellow one -- to an oriented state with the population spread over both even and odd $c_j$'s. 

Fig.~\ref{fig:2}(b) illustrates the dependence of the ground-state energy on $\alpha_2$ for $\alpha_1=0$ and $\alpha_1=1$, while Figs.~\ref{fig:2}(c) and (d) show the corresponding orientation cosine, $\langle \cos  \theta \rangle$, and alignment cosine, $\langle \cos^2 \theta \rangle$. For the case of $\alpha_1=0$, there occurs a localization transition in the vicinity of $\text{ln} [\alpha_2] = 3$, which corresponds to a transition from a spherically symmetric to an aligned impurity. As shown in Figs.~\ref{fig:2}(d), such an isotropic-to-aligned transition is characterized by a sudden change in the alignment cosine, $\langle \cos^2 \theta \rangle$.

At a finite value of $\alpha_1 = 1$, however, the behavior of the system changes. Around  $\text{ln} [\alpha_2] = 3$ there occurs an isotropic-to-oriented transition, accompanied by a sudden change in $\langle \cos  \theta \rangle$, see Figs.~\ref{fig:2}(c). However, once the parity-preserving coupling  approaches the value of $\text{ln} [\alpha_2] \approx 5$, a smooth crossover to the aligned phase occurs. This crossover,  marked in Figs.~\ref{fig:2}(a) by a dash-dotted line,  is not accompanied by a change in the derivative of the ground-state energy and therefore does not represent a sharp transition between the states of different symmetry.

\section{Conclusions and outlook}

In summary, we have studied the delocalization-to-localization transition for a quantum rotor coupled to a bosonic bath. It was shown that, unlike in the polaron problem, the transition from a spherically-symmetric to a localized ground state occurs already at the mean-field level. Furthermore, depending on the symmetry of the interactions, the state can be oriented (broken parity) or aligned (definite parity), making it possible  to observe a crossover between the two symmetries in the localized phase.

Among the experimental systems to address the localization transition,   the most promising one is given by cold molecules trapped in superfluid helium nanodroplets~\cite{molecule_experiment_angular_superfluid_helium, LemeshkoDroplets16}. There, it is possible to trap slowly rotating molecules featuring highly-anisotropic interactions with helium. Moreover, angulon self-localization can potentially be studied in experiments on Rydberg excitations in BEC's~\cite{BalewskiNature13}, where orbital-angular-momentum-changing collisions between the Rydberg electron and ultracold atoms have already been observed~\cite{SchlagmullerPRX16}. Finally, studies of coupling between rotations and vibrations have a long history in the context of finite systems, such as nonspherical atomic nuclei~\cite{RoweWoodNuclearModels}, flexible polyatomic molecules~\cite{SchmiedtPRL16}, and electron bubbles in superfluid helium~\cite{TempereEPJB, VadakkumbattNatComm14}. Recasting these problems in terms of the angulon quasiparticle might give further insights into the angular localization transition discussed in this paper.

It is worth noting that here we undertook only the first step in the studies of self-localization of quantum rotors. For Fr\"ohlich polarons it has been demonstrated that a sharp self-trapping transition arises as an artefact of the mean-field approximation, since mean field favours symmetry breaking even if it is prevented by quantum fluctuations~\cite{FisherPRB86, no_self_trapping1, no_self_trapping2, no_self_trapping_numerical, Feranchuk05}. Thus, it still remains to investigate whether such a transition actually takes place for rotating impurities. Therefore, in order to get  a deeper understanding of angular self-localization, approaches beyond mean field need to be developed for the angulon problem. We hope that the results presented here will stimulate future studies of angular self localization, both in the context of the Hamiltonian~\eqref{eq:1}, and for extended angulon models including non-linear coupling terms.  Finally, studying an ensemble of interacting quantum rotors in a superfluid might pave the way to studying new phenomena related to quantum glassiness~\cite{YePRL93} and many-body localization~\cite{NandkishoreMBLreview}.

\acknowledgements
 RS acknowledges support by the European Research Council (ERC) under the European Union's Horizon 2020 research and innovation programme (grant agreement No 694227), and by the Austrian Science Fund (FWF), Project No. P27533-N27. ML acknowledges support by the Austrian Science Fund (FWF), Project No. P29902-N27.

\bibliography{Self_Localized}
\end{document}